\documentclass[12pt,letterpaper]{article}

\usepackage{paper2e}
\usepackage{mydefs2e}
\usepackage{xspace}
\usepackage{epsfig}

\newcommand{\Sla}[1]%
{\kern0.12em{\raise.15ex\hbox{$/$}\kern-.74em #1}}

\newcommand{\LEW}{\Lambda_{\rm EW}}
\newcommand{\LUV}{\Lambda_{\rm UV}}

\renewcommand{\d}{\partial}

\newcommand{\nn}{\nonumber}
\newcommand{\order}[1]{\scr{O}\left(#1\right)}

\newcommand{\op}{\scr{O}}

\begin{document}

\begin{titlepage}
\preprint{BUHEP-04-12\\
UMD-PP-05-017}

\title{Conformal Technicolor}

\author{Markus A. Luty$\,^{\rm a,b,c}$\ and\ Takemichi Okui$\,^{\rm b}$}

\address{$^{\rm a}$Physics Department, University of Maryland\\
College Park, Maryland 20742}

\address{$^{\rm b}$Physics Department, Boston University\\
Boston, Massachusetts 02215}

\address{$^{\rm c}$Jefferson Laboratory of Physics, Harvard University\\
Cambridge, Massachusetts 02138}

\begin{abstract}
We point out that the flavor problem in theories with dynamical
electroweak symmetry breaking can be effectively decoupled if the physics above
the TeV scale is strongly conformal, and the electroweak order parameter
has a scaling dimension $d = 1 + \ep$ with $\ep \simeq 1/{\rm few}$.
There are many restrictions on small values of $\ep$:
for $\ep \ll 1$, electroweak symmetry breaking requires a fine-tuning
similar to that of the standard model;
large-$N$ conformal field
theories (including those obtained from the AdS/CFT correspondence)
require fine-tuning for $d < 2$;
`walking technicolor' theories cannot have $d < 2$, according to
gap equation analyses.
However, strong small-$N$ conformal field
theories with $\ep \simeq 1/{\rm few}$ avoid all
these constraints, and 
can give rise to natural dynamical electroweak symmetry breaking 
with a top quark flavor scale of order $10^{1/\ep}\TeV$, 
large enough to decouple flavor.
Small-$N$ theories also have an acceptably small Peskin-Takeuchi
$S$ parameter.
This class of theories provides
a new direction for dynamical electroweak symmetry breaking
without problems from flavor or electroweak precision tests.
A possible signal for these theories is a 
prominent scalar resonance below the
TeV scale with couplings similar to a heavy standard model Higgs.
\end{abstract}

\end{titlepage}

\section{Introduction}
\label{sec:intro}
How is electroweak symmetry broken?
The most important theoretical clue we have is the hierarchy problem,
the problem of understanding the smallness of the weak scale compared
to much higher scales in physics such as the Planck scale.
Perhaps the most elegant solution of the hierarchy problem is
dynamical electroweak symmetry breaking \cite{TC}.
This is the idea that the scale of electroweak symmetry breaking is
determined by a new strong interaction scale.
This naturally explains the smallness of the electroweak scale, since
the strong interaction scale is given in terms of UV quantities by
\beq[hierarchy]
\LEW \sim \LUV \, e^{-g_{\rm c}^2/g_{\rm UV}^2},
\eeq
where $g_{\rm UV}$ is the strength of the coupling in the UV
and $g_{\rm c} \sim 4\pi$ is the critical value where electroweak symmetry
is broken.
For $g_{\rm UV} < g_{\rm c}$, the electroweak scale is naturally exponentially
small compared to $\La_{\rm UV}$.
This mechanism is already realized in nature in the strong interaction
sector, explaining why the QCD scale is naturally small compared to
higher scales.

This paradigm for electroweak symmetry breaking makes the general prediction
that the electroweak symmetry breaking sector is strongly coupled at the
TeV scale.
Within a few years, the LHC will definitively settle the fundamental
question of whether electroweak
symmetry breaking sector is weakly or strongly coupled.

Until the LHC turns on, we must rely on indirect constraints.
Dynamical electroweak symmetry breaking
faces a number of potential difficulties.
First, strong interactions at the TeV scale
can ruin the agreement of the
standard model with precision electroweak data.
However, if the physics that breaks electroweak
symmetry is a strongly coupled theory with no large or small
parameters, `\naive dimensional analysis' (NDA) gives an estimate for the
Peskin-Takeuchi $S$ and $T$ parameters
\beq[STfromNDA]
S_{\rm NDA} \sim \frac{1}{\pi},
\qquad
T_{\rm NDA} \sim \frac{1}{4\pi}.
\eeq
For comparison,
the value of the $S$ parameter from scaled-up QCD is \cite{SQCD}
\beq
S_{\rm QCD} \sim 0.3.
\eeq
These are rough estimates,
and are comparable to the size of the current 95\% confidence
level bounds \cite{STbound}.
These do not rule out models of dynamical
electroweak symmetry breaking.
The models that \emph{are} ruled out (without fine tuning) are
those containing a large number $N$ of degrees of freedom,
in which $S \sim N / \pi$.
These include large `technicolor' or `walking technicolor' 
theories \cite{WTC},
and Randall-Sundrum (RS) models \cite{RS} with gauge fields in the 
bulk \cite{bulkgauge}, 
which are related to large-$N$ conformal theories (CFT's)
by the AdS/CFT correspondence \cite{AdSCFT}.

Another general problem with models of dynamical electroweak symmetry
breaking is that flavor is generally not decoupled from the TeV scale.
In technicolor models, this is because the order parameter
that breaks electroweak symmetry is a techni-fermion bilinear $\bar\psi \psi$
with mass dimension $d = 3$.
The standard-model fermion masses therefore arise from 4-fermion operators
connecting the standard model fermions with the technifermions \cite{ETC}.
These operators have dimension 6, and therefore become strong at low scales.
In particular the top coupling becomes strong at a scale
\beq
\hbox{\rm QCD-like technicolor:}\ \ \ 
\La_t \sim \LEW \left(\frac{\LEW}{m_t}\right)^{1/2} \sim 5 \TeV,
\eeq
where $\LEW \sim 4 \pi v \sim 2 \TeV$ is the scale where the electroweak
symmetry breaking sector becomes strongly coupled.
$\La_t$ is the scale where flavor must be addressed in these models.

The flavor problem is less severe in models of `walking' technicolor,
in which it is assumed that the electroweak order parameter $\bar\psi \psi$
has a large anomalous dimension, and scales as an operator
with dimension $d = 3 - \ga$ \cite{WTC}.
Walking technicolor theories are similar to a CFT with a nearly marginal
(slightly relevant) operator
that runs slowly and becomes strong and breaks electroweak symmetry.
Analyses based on the truncated Schwinger-Dyson equations show that
in asymptotically free theories, $d \ge 2$ \cite{WTC,CG}.
The scale where the top coupling becomes strong is then raised for $d=2$ to
\beq
\hbox{\rm Walking technicolor:}\ \ \ 
\La_t \lsim \LEW \, \frac{\LEW}{m_t} \sim 10 \TeV.
\eeq
Attempts to make realistic models based on strong top dynamics can be found
in \Refs{strongtop}.
In this paper, we will instead attempt to avoid strong flavor-dependent
dynamics at low scales.

A simple way to avoid the restriction $d \ge 2$ is to assume that the theory
is at an interacting conformal fixed point above the TeV scale.
This class of theories offers a solution of the hierarchy problem that
is identical to asymptotically free theories such as technicolor.
If the CFT is coupled to a gauge theory that is asymptotically free,
this gauge theory will become strong in the IR, causing the CFT to
flow away from its fixed point.%
\footnote{The same mechanism was employed for walking technicolor theories in
\Ref{PMTC}, where the QCD gauge coupling plays the role of the asymptotically
free gauge group.
This mechanism na\"\i{}vely predicts $\La_{\rm EW} \sim \La_{\rm QCD}$,
and we do not consider it here.}
The resulting non-perturbative dynamics can give rise to electroweak
symmetry breaking.
Another possibility is that the CFT contains a nearly marginal operator
that becomes strong in the IR.
These mechanisms are attractive because it generates an exponentially
large hierarchy.
Another possible mechanism exists if the CFT has a relevant operator that
transforms nontrivially under a global symmetry, \eg a discrete symmetry.
The coefficient of this operator can then be naturally small, and can set
the scale for the breaking of conformal and electroweak symmetry.
In this mechanism, the large hierarchy is put in by hand in the form of a
small coefficient, but it is technically natural.

In a strong CFT, flavor arises from couplings of the form $\bar{q} q \scr{O}$,
where $q$ is a standard-model fermion and $\scr{O}$ is a CFT operator
with quantum numbers of the Higgs.
In order to decouple flavor, we would like to have the scaling dimension
$d$ of the operator $\scr{O}$ as small as possible.
In CFT's, bound on the scaling dimension of a scalar
operator is $d \ge 1$ \cite{mack}.
In the limit $d \to 1$, the scalar operator behaves as a weakly-coupled scalar
field, which is just the standard-model Higgs.
The theory is therefore fine-tuned and does not solve the hierarchy problem.
However, for $d = 1 + \ep$, with $\ep \simeq 1/\hbox{\rm few}$,
the top quark becomes strongly
coupled at the scale
\beq[topmassscaling]
\hbox{\rm Conformal dynamics:}\ \ \ 
\La_t \sim \LEW \left( \frac{\LEW}{m_t} \right)^{1/\ep}.
\eeq
This scale is {\it exponentially} large for small $\ep$, and therefore we
can plausibly have sufficiently large $\ep$ to avoid fine-tuning, while
decoupling the flavor to high scales.
How $\La_t$ must be to avoid flavor-changing neutral currents
depends on the nature of flavor violation at this scale.
The most pessimistic case imaginable is that there are
unsuppressed strong contributions to operators that
contribute to $K$--$\bar{K}$ mixing at the scale $\La_t$.
This requires $\La_t \sim 10^5\TeV$, which is obtained for
$\ep \simeq \frac 15$.
If we assume some suppression of flavor violation for the
lightest generation, we expect that the flavor scale can be significantly
lower.
For example, a single Yukawa suppression of four-fermion operators
contributing to $K$--$\bar{K}$ mixing lowers the flavor scale to
$\La_t \sim 3 \times 10^{3} \TeV$, which requires $\ep \simeq \frac 13$.
Such values of $\ep$ are definitely plausible.
For example, in F-theory constructions of AdS$_5$ duals, one finds scalar
operators with dimension 
$\frac 43$ and $\frac 65$ \cite{Maldacena}.
The possible application of non-supersymmetric CFT's with low-dimension scalar
operators to the electroweak
hierarchy problem was also discussed in \Ref{strasslerCFT}.

The Randall-Sundrum model gives an explicit example of a 4D CFT, and
has been extensively discussed as a solution of the hierarchy problem.
In this model, the Higgs is usually localized in the the IR brane 
to obtain a large hierarchy.
In this case the Higgs field can be thought of as a bulk field with 
a large mass, 
and in the corresponding 4D interpretation the electroweak order 
parameter has a large ($d > 4$) scaling dimension.%
\footnote{By analogy to `walking technicolor,' this can be
thought of as `speeding technicolor.'}
To obtain sufficiently large fermion masses, the fermions are put on the IR
brane or in the bulk \cite{bulkfermions}.
In 4D language, this corresponds to generating fermion masses by making them
mix with composite fermions so that they can feel the symmetry breaking
in the strong sector.
The mixing of the standard-model fermions with composite fermions
was considered previously in the context of QCD-like 
technicolor \cite{Kaplan}.
Theories of this type are interesting alternatives to our scenario.  As we 
discuss in the appendix, these theories generally have a potentially
viable region of the 
parameter space where $\De\rho$ is just at the current experimental bound  
while corrections to $Z\to b\bar{b}$ require fine-tuning at the $10\%$ level 
\cite{Hopkins}. 
However, we will pursue scenarios where the standard-model fermions are 
completely elementary, just like in conventinal technicolor theories.

It is a simple matter to modify the RS model to give the electroweak
order parameter a smaller dimension:
one simply puts a Higgs scalar in the bulk, and leaves the fermions on the
UV brane.
Electroweak symmetry is broken by a Higgs potential localized on the IR 
brane (ensuring that this is an IR effect) and the bulk Higgs field communicates
electroweak symmetry breaking to the fermions on the UV brane.
Taking the Higgs bulk mass parameter to be negative makes the dimension of the
Higgs operator in the 4D CFT description smaller.
However, we will show that scalar operators with $d < 2$ are fine-tuned in RS.
This can be traced to the fact that RS is a large-$N$ theory, 
and this fine-tuning is common to all large-$N$ theories.

We are therefore led to a rather dark corner of theory space:
non-supersymmetric 4D strongly-coupled conformal field theories with small $N$.
These can have a scalar operator with dimension $d=1+\epsilon$
with $\ep \sim 1/\hbox{\rm few}$, and can dynamically break electroweak
symmetry at the TeV scale while giving large fermion masses without
flavor-changing neutral currents.  Small-$N$ theories also have an acceptably
small $S$ parameter. Not much is known about the dynamics of such theories,
and so our discussion of these theories is necessarily speculative.
Above the TeV scale, the theory becomes conformally invariant, and the new
strong conformal dynamics can be directly tested in direct analogy with the
way QCD is tested at a high-energy $e^+ e^-$ collider.%
\footnote{See \Ref{cftee} for a supersymmetric example.}
However, even the LHC will be limited to exploring the lightest `hadrons'
of the CFT, and it is not possible to make rigorous
predictions for this regime.
In the case where $\ep$ arises from a (moderately) small parameter in
the fundamental theory, we argue that 
the theory contains a prominent
scalar resonance near (but somewhat below) the TeV scale, with couplings
similar to those of a heavy standard-model Higgs, but deviating from
the standard-model couplings by order $\ep$. 
This provides an interesting and well-motivated
signal to look for at the LHC, whose observation
would clearly motivate going to even higher energies in the future.

This paper is organized as follows.
In section 2, we review the constraints on the operator dimension $d$
in various types of known models
and argue that small-$N$ theories avoid all constraints
and can have $d < 2$ without fine tuning.
In section 3, we study the phenomenology of these theories,
focusing mainly on the possibility of a Higgs-like scalar resonance. 
Section 4 contains our conclusions.

\section{Scalar Operators with $1 < d < 2$}
\label{sec:scalarop}
As discussed in the introduction, an important question for models
of dynamical electroweak symmetry breaking based on conformal field
theory is the scaling dimension $d$
of the `Higgs' operator $\scr{O}$ that acts as the electroweak order parameter.
In order to decouple flavor, we would like to have $d$ as small
as possible, while avoiding fine-tuning.
In this section we review what is known about low-dimension scalar operators,
and argue that
theories with strong coupling and small $N$ can give operators with
$d < 2$ without fine tuning. 

\subsection{$d \simeq 1$}
General theorems of conformal field theory tell us that a scalar
operator $\scr{O}$ must have dimension $d \ge 1$ \cite{mack}.
Furthermore, an operator with dimension $d = 1$ is a
free field (meaning that correlation functions of $\scr{O}$
are the same as those of a free field).
It is therefore clear that for $d$ sufficiently close to 1, the theory is
equivalent to a CFT weakly coupled to a Higgs
field, which clearly does not solve the hierarchy problem.
In CFT language, this is because the operator
$\scr{O}^\dagger \scr{O}$ has dimension close to $2d$,
which is relevant for $d < 2$.
The existence of a relevant operator that cannot be forbidden by
symmetries means that the fixed point is not reached unless the
coefficient of the relevant operator is tuned.
This argument seems to suggest that the theory is fine-tuned for any
$d < 2$, but it is limited to weak coupling because we have assumed
that the dimension of $\scr{O}^\dagger \scr{O}$ is approximately $2d$.
Since the anomalous dimension $d-1$ is of order a one-loop factor, we expect 
the anomalous dimension of $\scr{O}^\dagger \scr{O}$ to also be of order 
$d - 1$ for a weakly-coupled theory with $d - 1 \ll 1$.
But for strong coupling the operator $\scr{O}^\dagger \scr{O}$ will have
a large anomalous dimension, and its dimension will not have any
simple relation to the dimension of $\scr{O}$.%
\footnote{In fact, for a strong CFT the operator ``$\scr{O}^\dagger \scr{O}$''
has no {\it a priori} meaning.
The remarks above apply if we define $\scr{O}^\dagger \scr{O}$
to be the operator of lowest scaling dimension in
the operator product expansion of $\scr{O} \times \scr{O}^\dagger$ other
than the unit operator.}
Exceptions to this are large-$N$ theories, as we will discuss below. 

How small can we take $\ep = d - 1$?
In a nearly free theory with an elementary scalar, $\ep$ is an
anomalous dimension, which is of order a loop-counting factor in the
theory.
In other words, perturbation theory is an expansion in powers of
$\ep$, and we expect it to break down when $\ep \gsim 1$.
For values like $\ep = 1/{\rm few}$, perturbation theory is no longer
a reliable guide.
In these theories, it is possible that the operator $\scr{O}^\dagger \scr{O}$
has a sufficiently large anomalous dimension to make it an irrelevant operator.
We conclude that the general theorem does not
imply that theories with $\ep \sim 1/{\rm few}$ are fine-tuned.

\subsection{Large-$N$ Theories}
We now consider large-$N$ CFT's.
In large-$N$ theories, the matrix elements of operators 
factorize, 
and we can conclude that the operator $\scr{O}^\dagger \scr{O}$ has
dimension
near $2d$ (up to $1/N$ corrections) even if the theory is strongly
coupled.  This is true even for theories with a large 't Hooft parameter, 
such as the 4D dual of the Randall-Sundrum (RS) model, as we will discuss 
below.
The operator $\scr{O}^\dagger \scr{O}$ is therefore relevant for
all $d < 2$,
and we conclude that large-$N$ CFT's with
scalar operators with $d < 2$ are fine-tuned.

Large-$N$ theories are also disfavored as a dynamical electroweak symmetry
breaking sector because they give contributions to the $S$ parameter
that grow with $N$.  In theories with a small 't Hooft parameter,
this is simply because $S$ arises from a vacuum polarization effects
that count the number of microscopic states.
We will see that RS models, which have a large 't Hooft parameter, also 
predict large $S$ in the absence of fine-tuning.

For these reasons, we are led to consider theories that do not
have large $N$, and which cannot be obtained from RS
setup.

\subsection{QCD-like Theories}
$SU(N)$ gauge theories with $F$ flavors of Dirac fermions loses asymptotic
freedom for $F/N > \frac{11}{2}$, and for $F/N = \frac{11}{2} - \de$
with $\de \ll 1$
the theory has a weakly coupled `Banks-Zaks' fixed point \cite{BZ}.
This allows us to infer the existence of QCD-like theories that
flow to strongly coupled CFT's in the IR.
The loop expansion parameter at the Banks-Zaks fixed point is $\de$,
so we know there is a conformal window for a range of $\de$.
The conformal window ends at some value $\de \sim 1$, and the CFT's
at this end of the conformal window are necessarily strongly coupled.
Although this argument is strictly speaking limited to large $N$
(where $\de$ can be thought of as a continuous parameter),
it is very reasonable to assume that there are also small-$N$
asymptotically free gauge theories that flow to strong conformal
fixed points.

It is therefore natural to ask whether theories of this kind can
give rise to a conformal sector with scalar operators with $d < 2$.
To obtain a tractable approximation to the non-perturbative dynamics
of QCD-like theories, it is traditional to truncate the Schwinger-Dyson
equations by replacing the full gauge
propagator and gauge-fermion vertices by their tree-level values in the 
Landau gauge, giving rise to the so-called `gap equation.' 
However, this limits the parameterization of the non-perturbative
effects to spontaneous chiral symmetry breaking via
a dynamically generated `constituent quark mass.'
In particular, it is not able to describe the conformal fixed point
dynamics of the Banks-Zaks fixed point.

The conformal window was studied using the gap equation in \Ref{conformgap},
which concluded that the conformal window in QCD ends at $N_f \simeq 4 N_c$.
It would be very interesting to further
explore non-perturbative approximations
to QCD-like theories that can capture the physics of the perturbative
end of the conformal window and model the non-perturbative dynamics
at the strong end of the conformal window.
This is beyond
the scope of the present work.

\subsection{$d$ in Randall-Sundrum Models}
\label{subsec:dRS}

We now consider the Randall-Sundrum (RS) model,
which gives us a concrete, calculable CFT.
To make the connection to conformal field theory as
transparent as possible, it is 
convenient to write the 5D AdS metric as
\beq
   ds^2 = \mu^2 \eta_{\mu\nu} dx^\mu dx^\nu - \frac{1}{k^2} \frac{d\mu^2}{\mu^2},
\eeq
where $k$ is the AdS curvature,
and $\mu_{\rm IR} \le \mu \le \mu_{\rm UV}$ 
parameterizes the extra dimension.
The fact that this metric is invariant under the transformation, 
$x^\mu \to s^{-1} x^\mu$ and $\mu \to s\mu$, naturally leads to the 
interpretation that the parameter $\mu$ is proportional to the energy scale in 
the corresponding 4D CFT.
We assume that all dimensionful quantities in the 5D lagrangian are $\order{1}$
in units of $k$.
The physical size of dimensionful couplings at a position $\mu$ in the bulk
is then given by $k\mu$.
For simplicity, we will use units where $k = 1$.

We consider a 5D complex
scalar doublet $H$ with bulk mass $M$.
We will be interested in the situation where $H$ gets a VEV, breaking
an $SU(2)_L$ gauge symmetry.%
\footnote{We will ignore $U(1)_Y$ in this section for simplicity.  We will 
also ignore an $SU(2)_R$ gauge symmetry we need to put in the bulk for 
any realistic model to guarantee a custodial $SU(2)$ symmetry in the 4D CFT.
Adding these complications is straightforward and does not alter our 
conclusions.}
It is therefore convenient to parameterize the field by
\beq[HPhiPidefn]
H = e^{i\Pi_a \tau_a} \left( \begin{array}{c} 0 \\ \Phi \end{array} \right),
\eeq
where $\Phi$ and $\Pi_a$ ($a = 1,2,3$) are real fields.
When $\Phi$ gets a VEV, the zero mode of $\Pi_a$ parameterizes the Goldstone
degrees of freedom.
In this section, we are interested in the effects of the VEV, and so
we will mainly concentrate on the field $\Phi$.
Solving the equation of motion in the bulk,  
we obtain the most general 4D Poincare invariant solution:
\beq[Phisoln]
\Phi = A \mu^{n - 4} + B \mu^{-n}
\eeq
where $n \equiv 2 + \sqrt{4 + M^2}$.  
We restrict attention to masses satisfying the 
Breitenlohner-Freedman bound $M^2 \ge -4$ \cite{BFbound}, so that $n \ge 2$.
For the special case $n = 2$, the general solution is
\beq[Phisolnmod]
\Phi = \tilde{A} \mu^{-2} + \tilde{B} \mu^{-2} \ln\mu.
\eeq
We will focus on the generic case \eq{Phisoln} below.

We now discuss the CFT interpretation of the solution \Eqs{Phisoln}
for $n > 2$.
According the the AdS/CFT correspondence, the field $\Phi$
is associated with an operator $\scr{O}$ of the 4D CFT.
As we will review below, in the conventional case the operator $\scr{O}$
has dimension $d = n$, which implies $d \ge 2$.
To obtain a 4D CFT with $d < 2$, we must change the UV boundary 
condition \cite{Witten}.
Specifically, we must choose the UV boundary condition to set $B = 0$,
so that the IR boundary condition determines $A$.
This can be done by adding a UV boundary lagrangian
\beq
\scr{L}_{\rm UV} = -m |H|^2 = -m \Phi^2 .
\eeq
The UV boundary condition is then
\beq
\left[ \mu \frac{\d\Phi}{\d\mu} + m \Phi \right]_{\rm UV} &= 0,
\\
\eql{PiUVbdy}
\left. \mu \frac{\d\Pi_a}{\d\mu} \right|_{\rm UV} &= 0,
\eeq
which gives
\beq[UVboundary]
(n - 4 + m) A \mu_{\rm UV}^{n - 4} = (n - m) B \mu_{\rm UV}^{-n}.
\eeq
The generic solution is $A/B \sim \mu_{\rm UV}^{4 - 2n}$, but we can
tune $B = 0$ by taking
\beq[tunedmass]
m = 4 - n.
\eeq
Note that we are tuning a relevant operator that cannot be forbidden by
symmetries.

We now specify the IR boundary conditions.
We assume that the IR boundary conditions set
\beq
\eql{PhiIRbdy}
\left. \Phi \right|_{\rm IR} &= \ka,
\\
\eql{PiIRbdy}
\left. \mu \frac{\d \Pi_a}{\d\mu} \right|_{\rm IR} &= 0.
\eeq
This can be viewed as the result of adding an $SU(2)$ invariant
IR potential of the form
\beq
V_{\rm IR} = \frac{\la}{2} (|H|^2 - \ka^2)^2
\eeq
and taking the limit $\la \to \infty$ so that the radial component of the 
field is frozen.
(Note that the boundary conditions \Eqs{PiUVbdy} and \eq{PiIRbdy}
ensure that the fields $\Pi_a$ each have a massless Goldstone zero mode.)
The results for the coefficients $B$ and $A$ are then
\beq
B = 
\begin{cases}
\ka \mu_{\rm IR}^n \left[ 1 + \scr{O} \left(
\left( \mu_{\rm IR} /\mu_{\rm UV} \right)^{2n - 4}
\right) \right]
& \ {\rm `generic'} \cr
0 \vphantom{\Bigl[}
& \ \,{\rm tuned} \cr
\end{cases}
\eeq
and
\beq
A =
\begin{cases}
\displaystyle
\frac{n - m}{n - 4 + m} B \mu_{\rm UV}^{4 - 2n}
& \ {\rm `generic'} \cr
\ka \mu_{\rm IR}^{4 - n}
 \vphantom{\Bigl[}
& \ \,{\rm tuned} \cr
\end{cases}
\eeq

Now suppose that the VEV for the field $\Phi$ spontaneously breaks an
$SU(2)$ gauge symmetry in the bulk
and generates a mass for a fermion localized on the UV brane.
This is the RS description of breaking the gauge symmetry by the 
VEV of the operator $\scr{O}$ and giving mass to an elementary fermion
coupling to $\scr{O}$.
The 4D gauge boson mass is
\beq[gaugemassRS]
m_W^2 &= \frac{g_4^2}{2} \int_{\mu_{\rm IR}}^{\mu_{\rm UV}} d\mu\,
\mu \, \Phi^2
\\
&=
\begin{cases}
\displaystyle\vphantom{\Biggl[}
\frac{g_4^2 \ka^2}{4(n - 1)} \mu_{\rm IR}^2 \left[
1 + \scr{O}\left( \left( \mu_{\rm IR} / \mu_{\rm UV} \right)^{2n - 4} \right)
\right]
& \ {\rm `generic'} \cr
\displaystyle\vphantom{\Biggl[}
\frac{g_4^2 \ka^2}{4(3 - n)} \mu_{\rm IR}^2 \left[
1 - \left( \mu_{\rm IR} / \mu_{\rm UV} \right)^{6 - 2n}
\right]
& \ \,{\rm tuned} \cr
\end{cases}
\eeq
where $g_4$ is the gauge coupling of the 4D gauge zero mode.
In the tuned case with $n > 3$, the gauge boson mass grows large as 
$\mu_{\rm UV}$ increases, so $m_W$ is sensitive to the UV scale.
The physical interpretation of this case is that the gauge
symmetry is broken in the UV, not in the IR. This means that this case does 
{\it not} describe dynamical symmetry breaking.  Therefore, we will restrict 
to $n<3$ in the rest of the paper.  However, note that even if $n<3$, 
taking the $n \to 3$ limit in the tuned case leads to
\beq[tunedmW]
m_W^2 \to \frac12 g_4^2\ka^2\mu_{\rm IR}^2 \ln\left( \mu_{\rm UV} / \mu_{\rm IR}
\right),
\eeq
so the gauge boson mass is logarithmically sensitive to the UV.
We will discuss the 4D interpretation of the logarithm shortly.

Now consider fermions, a doublet $Q$ and a singlet $t^c$ localized on 
the UV brane.
We add the UV boundary term
\beq[UVbdyferm]
\De\scr{L}_{\rm UV} = \mu_{\rm UV}^{-1}
\left[ \bar{Q} i \Sla{D} Q + \bar{t^c}\sla{\del}t^c \right] 
- \left[ y_t H Q t^c + \hc \right] ,
\eeq
where the factor $\mu_{\rm UV}^{-1}$ multiplying the kinetic term arises
from the conformal factor in the metric.
The fermion mass is therefore
\beq
m_t &= y_t \ka \mu_{\rm UV} \Phi_{\rm UV}
\\
&= 
\begin{cases}
\displaystyle\vphantom{\Biggl[}
\frac{2n - 4}{n - 4 + m}\,
y_t \ka \mu_{\rm IR}
\left( \mu_{\rm IR} / \mu_{\rm UV} \right)^{n - 1}
& \ {\rm `generic'} \cr
\displaystyle\vphantom{\Bigl[}
y_t \ka \mu_{\rm IR} \left( \mu_{\rm IR} / \mu_{\rm UV} \right)^{3 - n}
& \ \,{\rm tuned} \cr
\end{cases}
\eeq
We therefore obtain
\beq
\frac{m_t}{m_W} \sim
\begin{cases}
\displaystyle\vphantom{\Biggl[}
\left( \frac{\mu_{\rm IR}}{\mu_{\rm UV}} \right)^{n - 1}
& \ {\rm `generic'} \cr
\displaystyle\vphantom{\Biggl[}
\left( \frac{\mu_{\rm IR}}{\mu_{\rm UV}} \right)^{3 - n}
& \ \,{\rm tuned} \cr
\end{cases}
\eeq

In the 4D CFT, the UV boundary term \Eq{UVbdyferm} corresponds to coupling
an elementary fermion to the CFT operator $\scr{O}$:
\beq
\De\scr{L}_{\rm CFT} = c_t \scr{O} Q t^{\rm c} + \hc
\eeq
The results for the fermion and gauge boson masses are therefore
consistent with the 4D interpretation that
$\scr{O}$ has scaling dimension
\beq
d = 
\begin{cases}
n & \ {\rm `generic'} \cr
4 - n & \ \,{\rm tuned} \cr
\end{cases}
\eeq
In the tuned case, recall that we require $n < 3$ to avoid breaking the
gauge symmetry in the UV.
At the critical value $n = 3$ \Eq{tunedmW} shows that we have
logarithmic UV sensitivity.
This is easy to understand.
The effective lagrangian at the IR scale has the form
\beq
\scr{L}_{\rm eff} = Z |D_\mu h|^2 - V(h) 
+ \left[ y_t h Q t^c + \hc \right],
\eeq
where $h$ is the lightest KK mode of the scalar.
The logarithm in \Eq{tunedmW} reflects the fact that the wavefunction
factor $Z$ depends logarithmically on the UV scale,
because in the $n\to 3$ (\ie $d\to 1$) limit $h$ is just a weakly-coupled 
scalar.
(Note that $g$ and $\ka$ are quantities defined at the IR scale.)
There is no logarithm in $m_t$ as $n \to 3$ because the fermion Yukawa
coupling is not renormalized (apart from the wavefunction renormalization
in $Z$) within our approximation of treating $Qt^c$ as a 
background field.
We therefore conclude that the tuned case can indeed describe operators 
with dimension $1 < d < 2$.

Now we study how much fine tuning is actually necessary to reach the tuned 
case.
If we deviate from the fine-tuning condition \eq{tunedmass} by 
$\Delta m$, the coefficient $B$ now has to be non-zero to satisfy the UV 
boundary condition \eq{UVboundary}:
\beq
   B \sim \ka \, \Delta m \, \mu_{\rm IR}^n 
              \left( \frac{\mu_{\rm UV}}{\mu_{\rm IR}} \right)^{2n-4} . 
\eeq
The condition that 
this is a small correction at the IR brane requires 
$B \mu_{\rm IR}^{-n} \lsim \ka$, \ie 
\beq[tuneamount]
   \frac{\Delta m}{m} \lsim \left( \frac{\mu_{\rm IR}}{\mu_{\rm UV}} 
                            \right)^{4 - 2d} 
                      \ll 1   ,
\eeq
where we have used $n=4-d$.
This quantifies the amount of the fine-tuning.  Note that for $d=1$ we have 
the same tuning as in the standard model.  Of course, we need only something 
like $d = \frac 54$ to push the flavor scale up to $10^4$~TeV,
but even in this case 
the fine-tuning is $\sim 10^{-6}$, which is unacceptable.  

The 4D CFT interpretation of this fine-tuning is easy to understand.
Since the CFT contains an operator $\op$, it also contains the operator 
$\op^\dagger \op$ which is invariant under all symmetries.
The AdS/CFT correspondence relates operator products to
multi-particle states, and the dimensions of composite
operators factorize essentially
as a consequence of the factorization properties of multi-particle states
in a weakly-coupled 5D field theory.
The dimension of the operator $\op^\dagger \op$ is therefore $2d$,
which means it is a \emph{relevant} operator.
Note that \Eq{tuneamount} is exactly the amount of fine-tuning required
to suppress the effects of a relevant operator of dimension $2d$.

A convincing check of this interpretation of the fine-tuning
can be obtained by computing the vacuum energy in the presence of the VEV 
as a function of the tuning parameter $\De m$.
The bulk action integral vanishes thanks to the bulk equation 
of motion, leaving only the boundary terms.
Since the UV boundary condition 
is chosen to cancel the boundary term at the UV brane, we only get the 
contribution from the IR boundary:
\beq[eq:GWpot]
   V(\mu_{\rm IR}) 
      &\sim \mu^5 \Phi 
             \left. \frac{d\Phi}{d\mu} \right|_{\rm IR} \nn\\
      &\sim \mu_{\rm IR}^4 \left[ 1 + \frac{\Delta m}{m} 
                               \left( \frac{\mu_{\rm IR}}{\mu_{\rm UV}} 
                               \right)^{2d-4}
                          + \cdots \right]   . 
\eeq
On the other hand, in the 4D CFT language, we are adding the operator
\beq
   \Delta {\cal L}_{\rm UV} = \lambda_{\rm UV} \op^\dagger \op 
\eeq
in the UV.  If the dimension of $\op^\dagger \op$ is $D$, 
then the coupling $\lambda$ should scale with energy as
\beq
   \lambda (\mu) = \lambda_{\rm UV}
   \left( \frac{\mu}{\mu_{\rm UV}} \right)^{D-4}.
\eeq
The vacuum energy associated with $\op^\dagger\op$ therefore has 
the form
\beq
   V &\sim \mu_{\rm IR}^4 \left[ 1 + \lambda (\mu_{\rm IR})
                                    + \lambda^2 (\mu_{\rm IR})
                                    + \cdots \right]    \nn\\
     &\sim  \mu_{\rm IR}^4 \left[ 1 + \lambda_{\rm UV} 
                                     \left( \frac{\mu_{\rm IR}}{\mu_{\rm UV}} 
                                     \right)^{D -4}  + \cdots
                           \right].
\eeq
Comparing this with \Eq{eq:GWpot}, we see that $D$ is indeed equal to $2d$ 
and that the tuning $\De m \to 0$ precisely corresponds to setting 
$\la_{\rm UV} = 0$.  Note that this is a tree-level calculation 
on the AdS side, which corresponds to a calculation at the leading order 
in the $1/N$ expansion on the CFT side.
Loop corrections correspond to $1/N$ corrections,
and are expected to give corrections to the relation $D = 2d$.
These corrections are suppressed by at least a loop factor, and are
small whenever perturbation theory is under control.
They therefore cannot be used to make the RS model with $d < 2$ natural,
but it does illustrate that the relation $D = 2d$ is not exact in general.

We will have more to say about the 4D interpretation of the RS model
with $d < 2$ in subsection \ref{subsec:lightscalarRS} below.

\subsection{$S$ in Randall-Sundrum Models}
We showed above that $d < 2$ requires tuning in the RS setup,
but argued that this is because these are large-$N$ CFT's.
But there is another problem with large-$N$ theories.
The fact that RS theories are large-$N$ theories means that we also expect
them to give large contributions to the $S$ parameter.
However, because the 4D CFT corresponding to an RS model also has large
't Hooft parameter, we cannot use NDA to estimate $S$.
We therefore briefly review the size of $S$ in these theories.
Complete results with numerical coefficients can be found in \Refs{SinRS}.

Contributions to $S$ can arise in various ways, but here let us focus 
on the bulk contribution to $S$ arising from the
mixing between the unperturbed gauge zero mode and the bulk KK modes
via the Higgs kinetic term \cite{Hopkins}.  This is not necessarily
the largest contribution to $S$, but it can be easily estimated, and we will see
below that it is already too large.  The leading effect of this kind comes from 
tree-level mixing between the zero-mode gauge bosons and the excited KK modes:
\beq
S_{\rm bulk} \sim 16 \pi g_5^2 \left( \frac{v}{m_{\rm KK}} \right)^4,
\eeq
where $g_5$ is the 5D gauge coupling,
$v = \ka \mu_{\rm IR}$ is the 4D VEV that breaks electroweak symmetry,
and $m_{\rm KK} \sim \mu_{\rm IR}$ is the mass of the lightest KK mode.
(Recall we are using units where $k = 1$.)
The 5D gauge coupling parameterizes the CFT contribution to the running of
the 4D gauge coupling.
For example, if there is a gauge kinetic term localized on the UV brane
with coefficient $1/g_{\rm UV}^2$, we have
\beq
\frac{1}{g_4^2} = \frac{1}{g_{\rm UV}^2}
+ \frac{1}{g_5^2} \ln\frac{\mu_{\rm UV}}{\mu_{\rm IR}}.
\eeq
In order to avoid a Landau pole below the UV scale, we require
\beq
g_5^2 \gsim g_4^2 \ln(\mu_{\rm UV} / \mu_{\rm IR}).
\eeq
In order to have $S_{\rm bulk} \lsim S_{\rm NDA} \sim 1/\pi$ (roughly the
current experimental limit), we must have
\beq
m_{\rm KK}^2 \gsim 4 \pi v^2
\left[ g_4^2 \ln(\mu_{\rm UV} / \mu_{\rm IR}) \right]^{1/2}.
\eeq
This shows that small values of $S$ can only arise from a hierarchy
between $v$ and $m_{\rm KK}$, which however requires fine tuning.
The amount of fine tuning is given by
\beq
{\rm tuning} \sim \left( \frac{4 \pi v}{\La_{\rm IR}} \right)^2
\lsim \frac{1}{N_{\rm KK}^2} \,
\frac{4\pi}{\left[ g_4^2 \ln(\mu_{\rm UV} / \mu_{\rm IR}) \right]^{1/2}}
\eeq
where $\La_{\rm IR}$ is the cutoff in IR units and
$N_{\rm KK} = \La_{\rm IR} / m_{\rm KK}$ counts the number of KK
modes below the cutoff.
We see that we can make $S$ small only at the price of either
fine-tuning, or taking  $N_{\rm KK}$ to be small,
meaning that the cutoff is so low
that it describes only a small number of KK modes.
In this case, the extra dimension is not really buying any
predictive power relative to a general effective theory with
a cutoff near the TeV scale.
Said another way, since we must allow arbitrary
higher-dimension operators
suppressed by powers of $\La_{\rm IR}$,
the expansion parameter of the effective theory is
$1/N_{\rm KK}$.

\section{A Composite Higgs?}
\label{sec:pheno}
In this section, we discuss the general phenomenology of a electroweak
symmetry breaking sector consisting of a small-$N$ strongly coupled CFT
with an electroweak order parameter with dimension $d = 1 + \ep$, with
$\ep \sim 1/{\rm few}$.
In the case where the (moderately) small value of $\ep$ arises from a 
small parameter in the fundamental theory, we
argue that the theory contains a prominent sub-TeV scalar resonance 
which can be thought of as a Higgs that 
is `partially' composite at the TeV scale.
The compositeness is `partial' in the sense that the couplings 
of the scalar to the strong TeV scale dynamics 
is parametrically suppressed by powers of $\ep$.

\subsection{Light Scalars in the Randall-Sundrum Model}
\label{subsec:lightscalarRS}
Although we are interested in 4D CFT's that do not have a simple
higher-dimensional interpretation, we begin by determining the
properties of the lightest scalar KK modes in the RS model.
We find the results illuminating, and will argue that they have a 
simple physical interpretation that extends to the case of a
partially composite Higgs.

To make a fully realistic RS model, we need a model with custodial
symmetry.
As explained in \Ref{Hopkins} this can be done by gauging a
$SU(2)_L \times SU(2)_R$ in the bulk
and breaking this down to $SU(2)_L \times U(1)_Y$ on the UV brane.%
\footnote{\Ref{Hopkins} gauged
$SU(2)_L \times SU(2)_R \times U(1)_{B-L}$ in the bulk,
which is necessary to properly embed $U(1)_Y$ for bulk fermions.
If fermions are elementary, as assumed here, there is no need for the
$U(1)_{B - L}$ factor in the bulk gauge group.}
Since we are using RS only as a guide, we 
will ignore the issue of custodial symmetry and consider the model with
only $SU(2)_L \times U(1)_Y$ gauged in the bulk.
However, it is straightforward to construct a fully realistic RS model
at the price of fine-tuning.

We now consider the KK decomposition of the bulk scalar doublet $H$.
We are mainly interested in the KK modes of $\Phi$, which contains the 
light Higgs-like boson (see \Eq{HPhiPidefn}).
The eigenvalue condition for KK masses of the $\Phi$ with the fine-tuned
boundary condition \Eq{tunedmass} in the UV and the shifted boundary 
condition $\Phi(\mu_{\rm IR})=0$ in the IR which corresponds to \Eq{PhiIRbdy} 
after subtracting the VEV is given by
\beq[scalareigen]
\frac{J_{-1+\ep}(x)}{J_{1-\ep}(x)} = -\frac{J_{\ep}(y)}{J_{-\ep}(y)},
\eeq
where
\beq
x = \frac{m}{\mu_{\rm IR}},
\qquad
y = \frac{m}{\mu_{\rm UV}}.
\eeq
Expanding this for small $x$, $y$, and $\ep$, we obtain the mass of the 
lightest mode:
\beq
\eql{massform}
m_0^2 &= 4 \ep \mu_{\rm IR}^2 \left[
1 - \left( \frac{\mu_{\rm IR}}{\mu_{\rm UV}} \right)^{2\ep} \right]^{-1}
\Bigl[ 1 + O(\ep) + O(x^2) + O(y^2) \Bigr]
\\
\eql{scalarmasscase}
&\simeq
\begin{cases}
\vphantom{\Bigl]}
4 \ep \mu_{\rm IR}^2
& for $\ep \gg 1/\ln(\mu_{\rm UV}/\mu_{\rm IR})$
\\
\displaystyle\vphantom{\Biggl]}
\frac{2 \mu_{\rm IR}^2}{\ln (\mu_{\rm UV}/\mu_{\rm IR})}
& for $\ep \ll 1/\ln(\mu_{\rm UV}/\mu_{\rm IR})$
\\
\end{cases}
\eeq
The 4D CFT interpretation of these results is the following.
The scalar would be exactly massless in the limit of unbroken conformal
symmetry, but the conformal symmetry is broken both in the UV and
the IR.
For {\it finite} $\ep$ in the limit $\mu_{\rm UV} \to \infty$,
the IR breaking of conformal invariance dominates.  Hence, if the scalar 
couples to the CFT with full strength, the conformal breaking should give 
$m_0^2 \sim \mu_{\rm IR}^2$.  However, in our case we expect from the 
general CFT theorem that the scalar should
decouple from the CFT for $\ep \to 0$.  We exactly see this behavior 
in the first case in \Eq{scalarmasscase}, where we
have $m^2 \sim \ep \mu_{\rm IR}^2$, and the interpretation of this is that 
$\ep$ controls 
the couplings of the scalar to the part of the CFT in which conformal 
symmetry is spontaneously broken (by the IR brane).
On the other hand,
the UV breaking of conformal symmetry dominates for finite
$\mu_{\rm UV}$ and $\ep \to 0$.
In this limit the scalar decouples from the CFT at low energies
(as required by the general theorem) but only logarithmically, just like
an elementary scalar.

For realistic models, the question of whether the breaking of conformal
invariance is dominantly in the UV or the IR depends on the scale
$\mu_{\rm UV}$ where the theory approaches the conformal fixed point.
This scale could be as high as the Planck scale even if the flavor scale
is much lower, since the top flavor interactions may be a weak perturbation
on the strong CFT dynamics.
In this case $\ln(\mu_{\rm UV} / \mu_{\rm IR}) \sim 40$, and since we 
are interested in $\ep \sim 1/{\rm few}$, we expect that
the breaking of conformal symmetry is dominated in the IR.
We will therefore mainly focus on this case in the following.

\begin{figure}[t]
\begin{center}
\includegraphics[width=12cm]{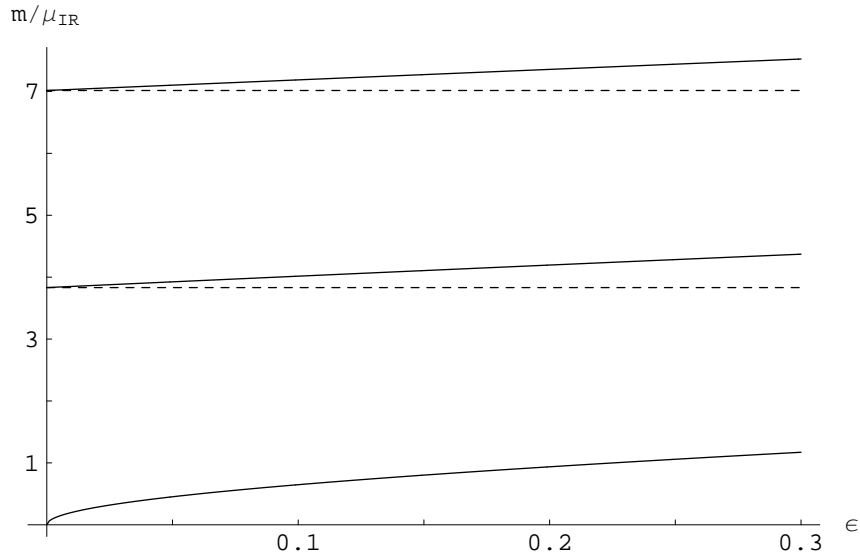}
\caption{The masses of the scalar (solid lines) and Goldstone (dashed lines) 
resonances as a function of $\ep$ in the RS version of the CFT.}
\label{fig:KKmasses}
\end{center}
\end{figure}

We have taken $\ep \ll 1$ in the above discussion in order to focus on the
parametrics, but we are really interested in $\ep \sim 1/{\rm few}$.
We can easily compute the mass of the scalar KK modes in this simple
model to see what we might expect for larger values of $\ep$.
In order to focus on the contribution to the scalar mass from the IR
breaking of conformal symmetry we take $\mu_{\rm UV} \to \infty$,
in which case the eigenvalue equation \Eq{scalareigen} for $\phi$
becomes
\beq[PhiKKeigen]
J_{-1 + \ep}(x)  = 0,
\eeq
while the eigenvalue equation for the `Goldstone' modes $\Pi_a$ is
\beq
J_1(x) = 0.
\eeq
These can be solved numerically, and the mass of the lightest
scalar resonances as a function of $\ep$ are plotted in
Fig.~\ref{fig:KKmasses}.
We see that there is a light `Higgs' even for moderate values of
$\ep$.  Note that for $\ep=0$ we recover a tower of completely degenerate 
doublets.%
\footnote{This is an exact statement since $J_{-1}(x)=-J_1(x)$.}

To get more evidences for our claim that $\ep$ controls the coupling of 
the scalar to the CFT, we now compute the couplings of this scalar resonance
to gauge bosons and fermions.
Writing the scalar field as
\beq
\Phi(x, \mu) = \sum_{n = 0}^\infty f_n(\mu) \phi_n(x),
\eeq
the KK wavefunctions $f_n$ are given by
\beq
f_n(\mu) = \frac{1}{\mu^2} \left[ A_n J_{-1 + \ep}\left( \frac{m_n}{\mu} \right)
+ B_n J_{1 - \ep} \left( \frac{m_n}{\mu} \right) \right].
\eeq
The Bessel functions are linearly independent for $0 < \ep < 1$,
which is sufficient for our purposes.
The boundary conditions determine the ratio $B_n/A_n$ and the mass $m_n$.
The UV boundary conditions then give
\beq
\frac{B_n}{A_n} = \frac{J_{\ep}(y_n)}{J_{-\ep}(y_n)}
\simeq \left( \frac{y_n}{2} \right)^{2\ep}.
\eeq
where $y_n = m_n / \mu_{\rm UV}$.
For simplicity we again take $\mu_{\rm UV} \to \infty$ to focus
on the IR effects, in which case this simply gives $B_n = 0$.

For the lightest KK mode we have $m_0 \ll \mu$ for all $\mu$, 
and we can expand the Bessel function $J_{-1+\ep}(y_0)$ to obtain
\beq[zeroKKwavefcn]
f_0(\mu) \simeq \frac{A_0}{\mu^2} \left[
\ep \left( \frac{m_0}{2\mu} \right)^{-1 + \ep}
- \left( \frac{m_0}{2\mu} \right)^{1 + \ep}
\right].
\eeq
The two terms are parametrically the same size at the IR brane because
$m_0^2 \simeq 4 \ep \mu_{\rm IR}^2$ (See \Eq{scalarmasscase}).
We determine $A_0$ by demanding that the real scalar field $\phi_0$ is
canonically normalized:
\beq
\sfrac 12 = \int_{\mu_{\rm IR}}^{\mu_{\rm UV}} d\mu\,
\mu\, f^2_0(\mu).
\eeq
The integral over the first term in \Eq{zeroKKwavefcn} involves
$\int d\mu\, \mu^{-1-2\ep} = -\mu / 2\ep$, and the additional
factor of $1/\ep$ makes this term dominate in the determination of $A_0$.
A similar integral appears when computing the $\phi W W$ coupling,
and we therefore find that the $\phi W W$ coupling is equal to the
standard model value up to $O(\ep)$ corrections.
More specifically, we find
\beq
\eql{phiWWinRS}
g_{\phi W W} = g_{\phi W W}^{\rm (SM)} \left[ 1 - \frac{\ep}{4} 
+ O(\ep^2) \right].
\eeq
Similarly, when computing the coupling of $\phi$ to the top quark,
we note that the first term in \Eq{zeroKKwavefcn} dominates at
large $\mu$, so the top coupling is also approximately equal to
the standard model, but this time with corrections suppressed by
$m_0^2 / \mu_{\rm UV}^2$.
This is so small that dominant corrections in fact come from 
perturbative top loops. Therefore, 
we see that in the RS model the couplings of the light scalar to top
quarks is very nearly equal to the standard model value, while
the couplings to gauge bosons has a larger, order $\scr{\ep}$ deviation 
from the standard model value.

We now describe the 4D CFT interpretation of these results.
The KK modes get massive due to the breaking of conformal
symmetry, which occurs in both the UV and the IR.
The UV breaking can be decoupled by sending $\mu_{\rm UV} \to \infty$.
As $\ep \to 0$ a full scalar doublet is becoming massless, with
deviations from standard model couplings suppressed (at least)
by powers of $\ep$.
Therefore, this supports our interpretation that the scalar doublet is 
coupled to the strong CFT with a coupling that vanishes as a power of $\ep$.
In RS setup fine-tuning is required to decouple the UV breaking of conformal
invariance, so this is not a viable solution to the
hierarchy problem, but as explained in section \ref{sec:scalarop}
this fine-tuning need not be present in small-$N$ models, which
we turn to next.

\subsection{Parametrically Light Scalars from Strong Conformal Dynamics}
We now consider a strongly coupled small-$N$
CFT with an electroweak order parameter (`Higgs operator') $\scr{O}$ with
dimension $d = 1 + \ep$.
The key feature of this theory is that $\ep$ is small enough to generate
a large hierarchy between the flavor scale and the electroweak scale
(see \Eq{topmassscaling}), but at the same time large enough that the operator
$\scr{O}^\dagger \scr{O}$ (`Higgs mass term') is irrelevant.
We argued above that for $\ep \ll 1$, the dimension of $\scr{O}^\dagger \scr{O}$
is $2d + O(\ep)$, which is therefore relevant.
We are assuming that $\ep \sim 1/{\rm few}$, but the $O(\ep)$ corrections to the
dimension of $\scr{O}^\dagger \scr{O}$ are large enough to make the operator
irrelevant.

There are at least two ways one can imagine this coming about.
The first possibility is that there is no large or small parameter in the theory,
and the fact that $d$ is close to $1$ is simply a numerical accident.
In this case, we expect the low-energy effective theory below the scale
$\La$ to be a general strongly coupled NDA theory with no small parameters.
Another possibility is that there is in fact a moderately small parameter
in the fundamental theory that controls the size of $\ep$.
(For example, in the RS model this parameter is the mass of the bulk scalar.)
For $\ep \sim 1/{\rm few}$ it is expected that not all quantities will be
numerically small even if parametrically suppressed by a power of $\ep$.
We assume one such numerical accident that the anomalous dimension of 
$\scr{O}^\dagger \scr{O}$ is large.%
\footnote{We remind the reader that the hierarchy between the flavor scale
and the electroweak scale is exponentially sensitive to $\ep$, and therefore
is expected to be more robust than the smallness of power corrections.}
In this subsection, we will describe
the effective field theory below the TeV scale for
this theory.
We will start with the limit $\ep \ll 1$, and then extrapolate
to $\ep \sim 1/{\rm few}$.

As argued above,
for $\ep \ll 1$ the correlation functions of the operator $\scr{O}$
are approximated by the correlation function of an elementary scalar
doublet $h$.
In the range of energies where the theory is approximately conformal,
the theory can therefore be written in terms of a lagrangian where
$h$ is coupled to a strong sector:
\beq[LeffScalar]
\scr{L} = | D h|^2 - m^2 |h|^2 - \sfrac 14 \la |h|^4
+ c_1 ( \scr{O}_{1} h^\dagger + \hc )
+ \cdots,
\eeq
where $\scr{O}_{1}$ is an operator in the strongly-coupled
sector of the theory with the same quantum numbers as $h$.
We are interested in the case where the coupling to the strong sector is
sufficiently strong so that the operators $|h|^2$ and $|h|^4$ have
large anomalous dimensions, and are irrelevant.
This allows us to neglect the couplings $m^2$ and $\la$ in the 
conformal regime without fine tuning, and is the basic reason that this
class of theories solves the hierarchy problem.
In general, large anomalous dimensions for the operators $|h|^2$ and
$|h|^4$ imply that $\ep \sim 1$, where $\ep$ is defined as the anomalous
dimension of $h$ itself.
However, it is an assumption of our analysis that $\ep \sim 1/{\rm few}$ is 
still small enough that an expansion 
in powers of $\ep$ gives at least qualitatively correct results 
for observables.

In order for the coupling $c_1$ to be important near the fixed point of
the theory, the scaling dimension of the operator must be sufficiently
small (3 in the limit $\ep \ll 1$).
We must assume that such an operator exists in order for $h$ to have
sufficiently strong couplings to the strong sector of the CFT.
In order to make a contribution of order $\ep$ to the anomalous dimension
of the scalar field, we must have
\beq[c1]
c_1 \sim 4\pi \sqrt{\ep}\,.
\eeq
The powers of $4\pi$ are counted using NDA, which assumes that the theory
has no large or small parameters other than $\ep$. This is valid because 
our theory is by assumption a small-$N$ theory.%
\footnote{NDA does not hold in the RS model, but the argument for the 
powers of $\ep$ does apply.}

We can also consider couplings such as
\beq
\De \scr{L} = c_2 \scr{O}_2 |h|^2 + c_3 ( \scr{O}_3 h^2 + \hc )
+ \cdots
\eeq
However, since by assumption $|h|^2$ has scaling dimension greater than 4,
$c_2$ is irrelevant near the fixed point.
The coupling $c_3$ is only important at the fixed point if the $SU(2)_W$
triplet operator $\scr{O}_3$ has
a sufficiently low dimension (2 in the limit $\ep \ll 1$), but there is no
reason to expect that an operator 
of such low dimension exists in the strong sector of the CFT.
This obviously generalizes to coupling involving higher powers of $h$,
and we conclude that the coupling of a single power of $h$ to the
strong sector of the CFT will dominate at the fixed point.

Now suppose that the conformal symmetry is broken at an IR scale $\La$.
(We assume that the conformal fixed point is reached at very high
energies, and therefore neglect any UV breaking of conformal symmetry.)
For example, we can have a new non-abelian gauge group that gauges a
global symmetry of the CFT.
If the new gauge coupling is asymptotically free, it will get strong in
the IR and break the conformal symmetry, as discussed in the introduction.
The point of this is that the breaking of conformal symmetry occurs 
in the strong sector of the CFT, since the new gauge fields do not 
couple directly to $h$.  Therefore, $h$ learns the conformal breaking only 
via coupling to the CFT, and we would like to understand how $\ep$ controls
this communication.

Below the scale $\La$, the coupling $c_1$ in \Eq{LeffScalar} will
generate all possible interactions of the scalar field $h$.
In particular, it will generate the $|h|^2$ and $|h|^4$ terms, which
are no longer rendered irrelevant by the strong conformal dynamics.
The effective lagrangian below the scale $\La$ is then
\beq[Leffscalar]
\scr{L}_{\rm eff} =  
|D h|^2 
+ \frac{\La^4}{16\pi^2}\, \scr{F}\left( \frac{4 \pi \sqrt{\ep}\, h}{\La},\,
\frac{D_\mu}{\La} \right),
\eeq
where $\scr{F}$ is an order-1 function that parameterizes the effects of the
strong sector, and $D_\mu$ is the gauge covariant derivative
of the standard model.
The factors of $4\pi$ are put in according to NDA, which assumes that 
the theory is strongly coupled at the scale $\La$ with no large or small
parameters other than $\ep$.
Here we are again using the assumption that this is a small-$N$ theory.
Expanding out \Eq{Leffscalar}, we obtain
\beq
\scr{L}_{\rm eff} = |D h|^2 - m_h^2 |h|^2 
- \sfrac 14 \la |h|^4 + \cdots,
\eeq
with
\beq
m_h^2 \sim \ep \La^2,
\qquad
\la \sim 16\pi^2 \ep^2.
\eeq
Note that the result for the mass agrees with the RS calculation 
\eq{scalarmasscase}.
If $m_h^2 < 0$, electroweak symmetry will be spontaneously broken by
a VEV for $h$, and we obtain
\beq
m_W^2 = \sfrac 12 g_2^2 v^2,
\qquad
m_Z^2 = \sfrac 12 (g_1^2 + g_2^2) v^2
\eeq
from the $h$ kinetic term, where
\beq
\avg{h} = \pmatrix{0 \cr v \cr},
\eeq
with $v = 174 \GeV$ as in the standard model.
We therefore obtain
\beq
\La \sim 4 \pi \sqrt{\ep}\, v.
\eeq
The scale of new strong dynamics is parametrically below the scale
$\LEW \sim 4\pi v \sim 2\TeV$, but for $\ep \sim 1/{\rm few}$ there are
large uncertainties of the NDA estimates.%
\footnote{%
Note that it does not make sense to take $\ep$ to be smaller than 
the gauge loop contribution to the anomalous dimension, $\sim g_2^2/16\pi^2$. 
In fact, at this critical value we have $\La \sim g_2 v \sim m_W$.}

We now discuss the couplings of $h$ to the standard model gauge bosons.
Because $\avg{4\pi \sqrt{\ep}\, h / \La} \sim 1$, 
the function $\scr{F}$ in the effective lagrangian \Eq{Leffscalar}
contains all possible electroweak breaking couplings of $h$ and gauge bosons
with $O(1)$ coefficients in terms of $4\pi \sqrt{\ep}\, h / \La$ 
and $\La$.
For example, the leading correction to the $hWW$ coupling is 
\beq
\De\scr{L}_{\rm eff} \sim \frac{\La^4}{16\pi^2} 
\left( \frac{g_2 W_\mu}{\La} \right)^2
\frac{4\pi \sqrt{\ep}\, h}{\La} 
\sim \ep g_2^2 v W_\mu^2 h + \cdots.
\eeq
This gives an $O(\ep)$ correction to the coupling of the Higgs
to gauge bosons, as anticipated from the RS computation \eq{phiWWinRS}.

\begin{figure}[t]
\begin{center}
\includegraphics[width=5cm]{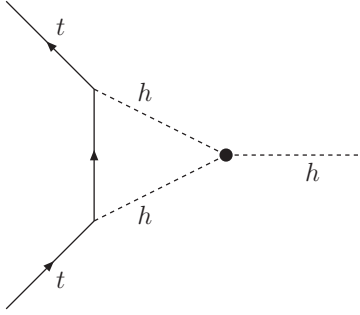}
\caption{Correction to the $ht\bar{t}$ coupling in the low energy effective
theory below $\La$.  The bullet represents a local operator that includes
CFT contributions to the 3-point function as well.} 
\label{fig:htt}
\end{center}
\end{figure}

We now consider the couplings to the top quark.
In the conformal regime, we add a coupling of the form
\beq
\De\scr{L} = c_t (h Q t^{\rm c} + \hc )
\eeq
between the scalar doublet $h$ and the elementary top quark.
The field $h$ has an anomalous dimension $\ep$ that suppresses
the coupling $c_t$ at low energies, so that the top quark Yukawa
coupling at the weak scale is
\beq
y_t = c_t(\La_t) \left( \frac{\La}{\La_t} \right)^{\ep},
\eeq
where $\La_t$ is the scale where the top quark coupling is generated.
As discussed above, we assume that $c_t(\La_t)$ is sufficiently large
that $y_t v = m_t$ at the weak scale.
The leading coupling of $h$ to the top quark is therefore the same as
in the standard model.
The leading corrections to this come from $h$ loop effects such as the
one shown in Fig.~\ref{fig:htt}, which gives
\beq
\De y_t \sim \frac{y_t^2}{16\pi^2} \frac{\La^4}{16\pi^2} 
             \left( \frac{4\pi\sqrt{\ep}\, h}{\La} \right)^3 
             \frac{1}{m_h}
\sim \frac{y_t^2}{4\pi} \ep,
\eeq
where the factor of $1/m_h$ arises because the diagram is IR dominated.
This diagram is the same as in the standard model, except that the 3 point
Higgs interaction does not have the standard model value.
This can be seen by expanding the function $\scr{F}$ in \Eq{Leffscalar},
where we see that the VEV and the 2- and 3-point functions of the Higgs
expanded about the VEV are independent.
This is expected to give a deviation from the standard model prediction
of between $1\%$ and $10\%$.

\subsection{Phenomenology}
We now turn to the phenomenology of this class of models.
One important question in these models is the size of precision electroweak
corrections.
The effective operator that contributes to $S$ is obtained
from the effective lagrangian \Eq{Leffscalar}:
\beq
\De\scr{L}_{\rm eff} \sim \frac{\ep g_1 g_2}{\La^2} 
h^\dagger W^{\mu\nu} h B_{\mu\nu}
\sim \frac{g_1 g_2}{16\pi^2} W^{\mu\nu}_3 B_{\mu\nu} + \cdots.
\eeq
This gives a contribution to $S$ with NDA strength, $S\sim 1/\pi$, \ie 
similar to small-$N$ technicolor.
The $T$ parameter requires breaking of custodial symmetry, and therefore
requires an additional hypercharge loop.
The effective coupling that gives rise to $T$ is therefore
\beq
\De\scr{L}_{\rm eff} \sim 
\frac{g_1^2}{16\pi^2}
\frac{16\pi^2 \ep^2}{\La^2} |h^\dagger D_\mu h|^2
\sim \frac{\ep g_1^2 m_Z^2}{16\pi^2} Z_\mu^2 + \cdots,
\eeq
where $Z_\mu$ is the $Z$ boson field.
This gives a contribution to $T$ that is parametrically smaller than the NDA
value:
\beq
T \sim \frac{\ep}{4\pi} \sim \ep \, T_{\rm NDA}.
\eeq
Taken at face value, this is not so
great for precision electroweak fits, since
the best fit for nonzero electroweak corrections has both $S$ and $T$
positive and comparable.
However, given the uncertainties in these estimates
these theories are still viable.

The phenomenology at high energy colliders depends on the structure of
strong resonances at the TeV scale.
We cannot make any rigorous prediction about these resonances other than
the fact that they must exist to unitarize WW scattering (the `no lose theorem').
In the case where the small dimension of the electroweak order parameter
is due to a moderately small parameter in the fundamental theory, we argued
above that there will be a prominent scalar resonance below the TeV scale
whose couplings to gauge bosons and the top quark are parametrically close
to the standard-model values.
The deviation from the standard-model value for the coupling to the top
quark is quite small, between $1\%$ and $10\%$, while the deviation from
the standard model value for the $\phi W W$ coupling is of order $\ep$,
which is expected to be at least $10\%$.
The possibility of studying a `non-standard' heavy Higgs-like scalar
at the LHC has been discussed by a number of authors \cite{Higgswidth}.
The discovery of such a particle
and the measurement of deviations from the standard-model predictions
for its couplings may well be the first indication of
strong conformal dynamics as the origin of electroweak symmetry breaking.

\section{Conclusions}
\label{sec:conc}
We have described a new paradigm for dynamical electroweak symmetry
breaking in which the electroweak symmetry breaking sector is a
conformal field theory (CFT) above the TeV scale. Conformal symmetry and
electroweak symmetry are broken at the TeV scale, triggered by an
asymptotically-free gauge group or a slightly relevant operator becoming 
strong, or by a relevant operator with a coefficient made small naturally 
by symmetries.  Any of these mechanisms stabilizes the weak scale against 
quantum corrections and gives a solution of the hierarchy problem.

The flavor problem of dynamical electroweak symmetry breaking is solved
if the dimension of the CFT operator that acts as the order parameter
for electroweak symmetry breaking has a dimension $d$ close to 1, the
dimension of an elementary Higgs scalar field. For $d = 1 + \ep$, the
scale where the top quark Yukawa coupling becomes strong is raised to
$\LEW (\LEW/m_t)^{1/\ep}$, where $\LEW \sim 2\TeV$. For $\ep \sim
1/{\rm few}$, this is large enough to effectively decouple flavor.

Finding a CFT with the required properties is highly nontrivial. Weakly
coupled CFT's can certainly have operators with dimension near 1 in the
form of an elementary scalar Higgs field $h$, but this clearly does not
solve the hierarchy problem because of the existence of the relevant
operator $h^\dagger h$ with dimension near 2. What is required is a
strongly-coupled CFT with a scalar operator $\scr{O}$ with
dimension $d = 1 + \ep$, where strong CFT dynamics renders the operator
$\scr{O}^\dagger \scr{O}$ irrelevant by giving it a large
anomalous dimension.  We have shown, however, that strong CFT's with
large $N$, including those obtained from the AdS/CFT correspondence,
have the property that the dimension of $\scr{O}^\dagger\scr{O}$ is
close to $2d$, and are therefore fine-tuned for $d < 2$.

We are therefore led to consider strongly-coupled, small-$N$ CFT's.
These theories also naturally have small electroweak precision
corrections, addressing another strong constraint on models of
dynamical electroweak symmetry breaking. The difficulty is that there
are no reliable theoretical tools for studying such theories, and in
fact no explicit examples are known. 
In the case where the smallness of $\ep$ is due to a small parameter
in the fundamental theory, we argued that there will be a prominent
scalar resonance with couplings to gauge bosons and
the top quark that are comparable to that of a heavy standard-model
Higgs, but with $O(\ep)$ deviations that can be measured at LHC.
We believe that this gives strong motivation to experimental studies of
a heavy Higgs-like particle, and look forward to a decisive test of
these ideas.

\section*{Acknowledgments}
We would like to thank N. Arkani-Hamed and A. Cohen for discussions, and
J. Terning and R. Sundrum for comments on the manuscript.  Finally, we 
express our gratitude to K. Agashe for pointing out an error in the appendix.
M.~A.~L. is supported by NSF grant PHY-009954.
T.~O. is supported by DOE contract DE-FG03-91ER-40676.

\startappendices
\setcounter{section}{0}

\section*{Appendix A: Composite Fermions}
\setcounter{section}{1}
In this appendix, we consider the possibility that fermion masses are
generated by mixing with fermionic operators of the CFT.
That is, we suppose that the CFT contains operators with quantum
numbers conjugate to the standard model fermion fields, and we include
interaction terms
\beq[flavormix]
\de\scr{L} = z_Q Q \scr{Q}^{\rm c} + z_u u^{\rm c} \scr{U}
+ z_d d^{\rm c} \scr{D}
+ z_L L \scr{L}^{\rm c} + z_e e^{\rm c} \scr{E},
\eeq
where $Q, \ldots, e^{\rm c}$ are standard model fermions, and
$\scr{Q}^{\rm c}, \ldots, \scr{E}$ are fermionic CFT operators.
The unitarity limit on the dimension of the CFT operators is $d = \frac 32$,
the dimension of a free fermion, so the couplings $z_Q, \ldots, z_e$ can
be marginal or even relevant without approaching the unitarity limit
for CFT operators.
This means that the flavor scale where these operators are generated
can be decoupled completely, just as in the standard model.
We will show that this mechanism generally require a mild $\sim 10$\% 
tuning to accomodate constraints on $Z \to b\bar{b}$ and the $T$ parameter 
together with the heavy top mass. 

This mechanism for generating fermion masses was first considered in
the context of technicolor theories in \Ref{Kaplan}. 
It has been revived recently in the context of RS theories, where it
corresponds to putting standard model fermions in the bulk \cite{bulkfermions}.
We consider this framework here using our CFT language, which makes 
it clear that our analysis is model-independent. 

In \Eq{flavormix}, the couplings $z_Q, \ldots, z_e$ are $3 \times 3$ matrices
in flavor space.
These $z$'s act as spurions that violate the $SU(3)^5$ flavor symmetry
that is otherwise present, and their transformation properties under the
flavor symmetry determine the structure of flavor violation in this model.
We will normalize the CFT operators $\scr{Q}^{\rm c}, \ldots, \scr{E}$ so
that $z \sim 1$ corresponds to strong coupling at the scale of conformal
and electroweak symmetry breaking $\La$.
If the conformal and electroweak symmetry breaking is strong with no large
or small parameters, NDA tells us that the quark mass matrices are
\beq
m_u = c \La z_Q^\dagger z_u,
\qquad
m_d = c \La z_Q^\dagger z_d,
\eeq
where $c \sim 1$.%
\footnote{We assume that the CFT preserves custodial symmetry so that the
coefficients $c$ are the same for up- and down-type fermions.}
To get the top quark mass, we must therefore have
\beq[topmassconstr]
(z_Q^\dagger z_u)_{33} \sim \frac{m_t}{\La} \sim \frac 1{10}.
\eeq

Because the $z$'s violate flavor and custodial symmetry, they give rise to
corrections to the $\rho$ parameter and the $Z \to b \bar{b}$ vertex.
Note that the leading corrections to the $\rho$ parameter do not involve 
internal gauge bosons, so we can ignore the custodial symmetry breaking
from $U(1)_Y$.
In this limit the standard model and the CFT each 
have a separate $SU(2)_L \times SU(2)_R$ symmetry, which are broken to 
diagonal subgroups by $z_Q$ and $z_{u,d}$.
Specifically, the $z$'s transform like
\beq
   z_L &\too L_{\rm SM} \, z_L L_{\rm CFT}^\dagger \nn\\
   z_R &\too R_{\rm SM} \, z_R R_{\rm CFT}^\dagger , 
\eeq
where $z_L \equiv z_Q$ and $z_R \equiv {\rm diag}(z_d, \, z_u)$.
Electroweak symmetry is broken with NDA strength in the strong sector
in the pattern $[SU(2)_L \times SU(2)_R]_{\rm CFT} \to SU(2)_{\rm cust, CFT}$,
so custodial symmetry breaking in the standard model sector depends
on the spurion $z_u z_u^\dagger - z_d z_d^\dagger$.
This spurion has $\De I_{\rm cust,SM} = 1$, while a
custodial symmetry violating contribution to the $Z$ mass
has $\De I_{\rm cust,SM} = 2$.
Therefore%
\footnote{The first version of this paper argued that
$\De M_Z^2 \sim z_u^\dagger z_u$.
We thank Kaustubh Agashe for pointing out our mistake.}
\beq
\De M_Z^2 \sim \frac{g^2 \La^2}{16\pi^2}
\tr(z_u z_u^\dagger - z_d z_d^\dagger )^2
\sim \frac{g^2 \La^2}{16\pi^2} \tr(z_u z_u^\dagger )^2
\eeq
In order to satisfy the constraint on the $\rho$ (or $T$) parameter, we require
\beq[rhoparamconstr]
\tr(z_u z_u^\dagger )^2  \lsim \frac 1{100}.
\eeq
On the other hand, the correction to the $Z \to b_L\bar{b}_L$ vertex is
\beq
\De g_{Z \to b\bar{b}} \sim 4\pi \frac{g}{4\pi} (z_Q^\dagger z_Q)_{33}.
\eeq
Satisfying the experimental constraint requires
\beq[Zbbconstr]
(z_Q^\dagger z_Q)_{33} \lsim \frac 1{100}.
\eeq

We see that generically there is a tension in satisfying all three constraints: 
the $\rho$ parameter constraint \eq{rhoparamconstr}, the $Z\to b\bar{b}$ 
constraint \eq{Zbbconstr} and the top mass condition \eq{topmassconstr}.
For example, with $(z_Q)_{33} \sim 1$, \Eq{topmassconstr} forces 
$(z_u)_{33} \sim \frac 1{10}$, which makes the $\rho$ constraint completely 
safe while requiring an additional 
contribution to the $Z \to b\bar{b}$ coupling that cancels the non-standard 
one to $1\%$ accuracy. With $z_u \sim 1$, the $Z\to b\bar{b}$ constraint is 
just on the edge while the $\rho$ constraint needs $1\%$ tuning.   
However, we can `compromise' and choose 
$(z_Q)_{33} \sim (z_u)_{33} \sim \frac 13$,
which puts us just on the edge of the $\rho$ constraint, while we must find 
additional contributions to $Z \to b\bar{b}$ that cancel the unwanted CFT 
contribution to $10\%$ accuracy.  Therefore, in this part of the parameter 
space, this framework is viable with mild tuning.

\end{document}